\documentclass[a4,aps,pre,twocolumn,nofootinbib,10pt]{revtex4-1}
\usepackage[latin1]{inputenc}
\usepackage{amsmath}
\usepackage{amsfonts}
\usepackage{amssymb}
\usepackage{graphicx,hyperref}
\usepackage{amssymb}
\usepackage{epstopdf,booktabs}
\usepackage{mathptmx}
\usepackage[usenames,dvipsnames]{color}
\graphicspath{{Figure/}}

\newcommand{\be}{\begin {equation}}
\newcommand{\ee}{\end {equation}}
\newcommand{\beqa}{\begin {eqnarray}}
\newcommand{\eeqa}{\end {eqnarray}}
\newcommand{\mb}{\mathbf}
\hypersetup{
    colorlinks,    citecolor=black,    filecolor=black,    linkcolor=black,    urlcolor=black
}

\begin{document}
\title{Thomson scattering in high-intensity chirped laser pulses}
 \author{Amol R. Holkundkar}
\email{amol.holkundkar@pilani.bits-pilani.ac.in}
\affiliation{Department of Physics, Birla Institute of Technology and Science, Pilani, Rajasthan, 333031, India}
\author{Chris Harvey}
\email{christopher.harvey@chalmers.se}
\affiliation{Department of Applied Physics, Chalmers University of Technology, SE-41296 Gothenburg, Sweden}
\author{Mattias Marklund}
\email[]{mattias.marklund@chalmers.se}
\affiliation{Department of Applied Physics, Chalmers University of Technology, SE-41296 Gothenburg, Sweden}

\begin{abstract}
We consider the Thomson scattering of an electron in an ultra-intense chirped laser pulse.  It is found that the introduction of a negative chirp means the electron enters a high frequency region of the field while it still has a large proportion of its original energy.  This results in a significant enhancement of the energy and intensity of the emitted radiation as compared to the case without chirping.

\end{abstract}

\maketitle

\section{Introduction}
 
In recent decades, since the discovery of chirped pulse amplification \cite{Strickland1985219}, the powers and intensities of laser facilities around the globe have been exponentially increasing \cite{PhysRevSTAB.5.031301}. The current record of $2\times 10^{22}$ W cm$^{-2}$ was set in 2008 \cite{Yanovsky:2008} and it is expected that this will be routinely surpassed in the near future at new facilities such as the Vulcan 20 PW upgrade \cite{Vulcan}, the Extreme Light Infrastructure (ELI) Facility \cite{ELI} and the XCELS project \cite{XCELS}.  The widespread availability of the current technology has driven a large field of research in nonlinear Thomson and Compton scattering, with a view to producing high-energy, tuneable $\gamma$-ray beams. These sources have important applications, both for fundamental research \cite{Wu} and for more practical applications, such as cancer radiotherapy \cite{Lawrence} and the radiography of dense objects \cite{PhysRevLett.94.025003}.  Recent experiments \cite{2011NatPh...7..867C, PhysRevLett.113.224801} have been pushing the limits of peak energies and brilliances.  However, working with ever-higher laser intensities we will soon enter a regime where radiation reaction (and, ultimately, QED effects \cite{Heinzl:2011ur, DiPiazza:2012RevModPhys}) will start to come into play.

Radiation reaction (RR) occurs when particles are accelerated so strongly by the laser field that their resulting radiation emissions cause significant energy losses.  The result is a frictional effect which can significantly impact on the particle dynamics, causing the particles to slow and reducing their energy as they reach the peak field \cite{Kravets:2013, Harvey:2011mp} (see also \cite{Heinzl:2015,Holkundkar:2014}).  As a consequence the resulting emission spectrum will be reduced in both energy and intensity (see, e.g. \cite{Hartemann:1996, Koga:2005}).  

Recently a number of articles have considered the effects of pulse chirping in laser-matter interactions.  This has been in the context of ion acceleration \cite{Yazdani:2014}, and in the Thomson scattering of relativistic electrons in moderately intense laser pulses \cite{Rykovanov:2014, Ghebregziabher:2013}.
In this paper we show how the introduction of a chirp into a very intense laser pulse can help to mitigate the reduction in energy of the emitted radiation by allowing the electrons to probe deeper into the laser focus before becoming susceptible to RR.  The result is a higher electron energy in the peak field, enabling a significant increase in the energies and brillances of the Thomson radiation.
 
\section{Theory}
We consider the interaction of a relativistic electron with a counter-propagating chirped Gaussian laser pulse of base frequency $\omega_0$ and FWHM duration $\tau_0$. We take the direction of the laser propagation to be along the $z$ axis, such that the laser wave vector is $\mb{k}=\omega_0\mb{\hat{z}}/c$.  Then we normalize space and time with respect to the wave vector and the base frequency, respectively ($x\rightarrow  k x$ and $t \rightarrow \omega_0 t$). Defining a chirped pulse in the same manner as a number of recent works (e.g.~Refs.~\cite{Yazdani:2014,Rykovanov:2014, Ghebregziabher:2013}), the vector potential can be written as
\begin{eqnarray}
\mb{A}&=& A_0 \exp\left( -\frac{\zeta\eta^2}{\tau_0^2}\right )\big[ \delta\cos (\eta+f(\eta))\mb{e}_x \nonumber\\
&&  +\sqrt{1-\delta^2}\sin (\eta+f(\eta))\mb{e}_y\big],
\label{eq:potential}
\end{eqnarray}
where $A_0$ is the field amplitude, $\zeta=4\ln(2)$, $\eta=t-z+\phi_0$, $\phi_0$ is a phase constant, and $f(\eta)$ is the chirp function\footnote{Note that introducing the chirp into the vector potential will result in different electromagnetic field components, and therefore different particle dynamics, to those obtained by introducing the chirp directly into the $\mb{E}$ and $\mb{B}$ fields.}.  
The factor $\delta$ determines the polarization, and is set to 1 ($1/\sqrt{2}$) for linear (circular) polarization.
The vector potential is then normalized as $\mb{A} \rightarrow e\mb{A}/m c$, and hence the electric field can be normalized as $\mb{E} \rightarrow e\mb{E}/m \omega_0 c$, where $\mb{A}$ and $\mb{E}$ are the vector potential and the corresponding electric field in SI units.   
 We then define the dimensionless intensity parameter $a_0=eE_0/m\omega_0 c$ in the usual manor, in terms of the peak fields.

A frequency chirped pulse is one where the frequency $\omega$ changes with time. This can be achieved by the use of a set of grating pairs \cite{Strickland1985447,kamada_chirp}, and it is now fairly trivial to introduce chirps such that the laser frequency changes by a few percent over the pulse duration.  In our expressions the chirping of the laser pulse is defined by the function $f(\eta )=b \eta^2$, where $b$ is the chirping constant.  The value of $b$ must be kept small since the since the chirp is restricted by the bandwidth, and in turn the length, of the initial generating pulse. Setting $b=0$ will correspond to the unchirped laser pulse.  It is important to note that the introduction of a chirp into the laser pulse will change the pulse energy ($\sim |\mb{E}|^2$). In order to compensate for this we reduce the pulse duration accordingly.  

Having described the pulse, we now turn our attention to the motion of the colliding electrons.
Ordinarily the particle motion would be governed by the Lorentz equation, but in cases where the acceleration is strong the emission of  radiation can lead to a significant reduction in the particle's energy and momentum.  
The effect of this `radiation reaction' on the particle dynamics can be included by adding a correctional term to the Lorentz force equation. 
However, determining what this correction should be is surprisingly  non-trivial. Despite having been studied for over 100 years, it remains one of the most fundamental problems in electrodynamics.
A common starting point is to solve the coupled Lorentz and Maxwell's equations for the system. 
This results in the infamous Lorentz-Abraham-Dirac equation \cite{Lorentz:1905,Abraham:1905,Dirac:1938nz}, which suffers from notorious defects such as pre-acceleration and (unphysical) runaway solutions.  
A common resolution to these problems is to adopt the perturbative approximation of Landau and Lifshitz \cite{LLII}.  Then the equation of motion is given by 
\begin{eqnarray}
\frac{d\mathbf{p}}{dt}=\mathbf{f}_\textrm{L}+\mathbf{f}_\textrm{R},
\end{eqnarray}
where $\mathbf{f}_\textrm{L}=\mathbf{E}+\mathbf{v}\times\mathbf{B}$ is the Lorentz force and the radiative correction term 
\begin{eqnarray}
\mathbf{f}_\textrm{R}&=&-\bigg( \frac{4}{3} \pi \frac{r_e}{\lambda}  \bigg)
\bigg\{\gamma\left[ \left( \frac{\partial}{\partial t}+\mathbf{v}\cdot\nabla\right)\mathbf{E}+\mathbf{v}\times \left( \frac{\partial}{\partial t}+\mathbf{v}\cdot\nabla\right)\mathbf{B}\right]\nonumber\\
&&+ \bigg[(\mathbf{f}_L)\times\mathbf{B}+(\mathbf{v}\cdot\mathbf{E})\mathbf{E}
- \gamma^2 [(\mathbf{f}_L)^2-(\mathbf{v}\cdot\mathbf{E})^2]\mathbf{v}\bigg]\bigg\},\label{LL}
\end{eqnarray}
where $r_e=e^2/mc^2$ is the classical electron radius and $\lambda=2\pi c/\omega_0$ is the base wavelength of the field.
Equation (\ref{LL}) is valid when the radiative reaction force is much less than the Lorentz force in the instantaneous rest frame of the particle.  We stress that there are a number of alternative equations in the literature (for an overview see \cite{Burton:2014,Vranic:2015}) and it is still an open problem as to which is the correct formulation.  However, it has recently been shown that the Landau-Lifshitz equation, along with some of the others, is consistent with quantum electrodynamics to the order of the fine-structure constant $\alpha$ \cite{0038-5670-34-3-A04,Ilderton:2013tb}.
Also, we note that the first term (derivative term) of Eq.~(\ref{LL}) is significantly smaller than the other two, since it is only linear in the field strength whereas the other terms are quadratic.  It is found that in almost all cases the contribution from this term is negligible and so we do not include it  in our simulations. (Indeed, it can be shown that, in cases where classical RR is important, the derivative term is even smaller than the electron spin force and so one could argue that it should be neglected out of consistency~\cite{Tamburini:2010}.)

\begin{figure}[t]
\includegraphics[width=1.0\columnwidth]{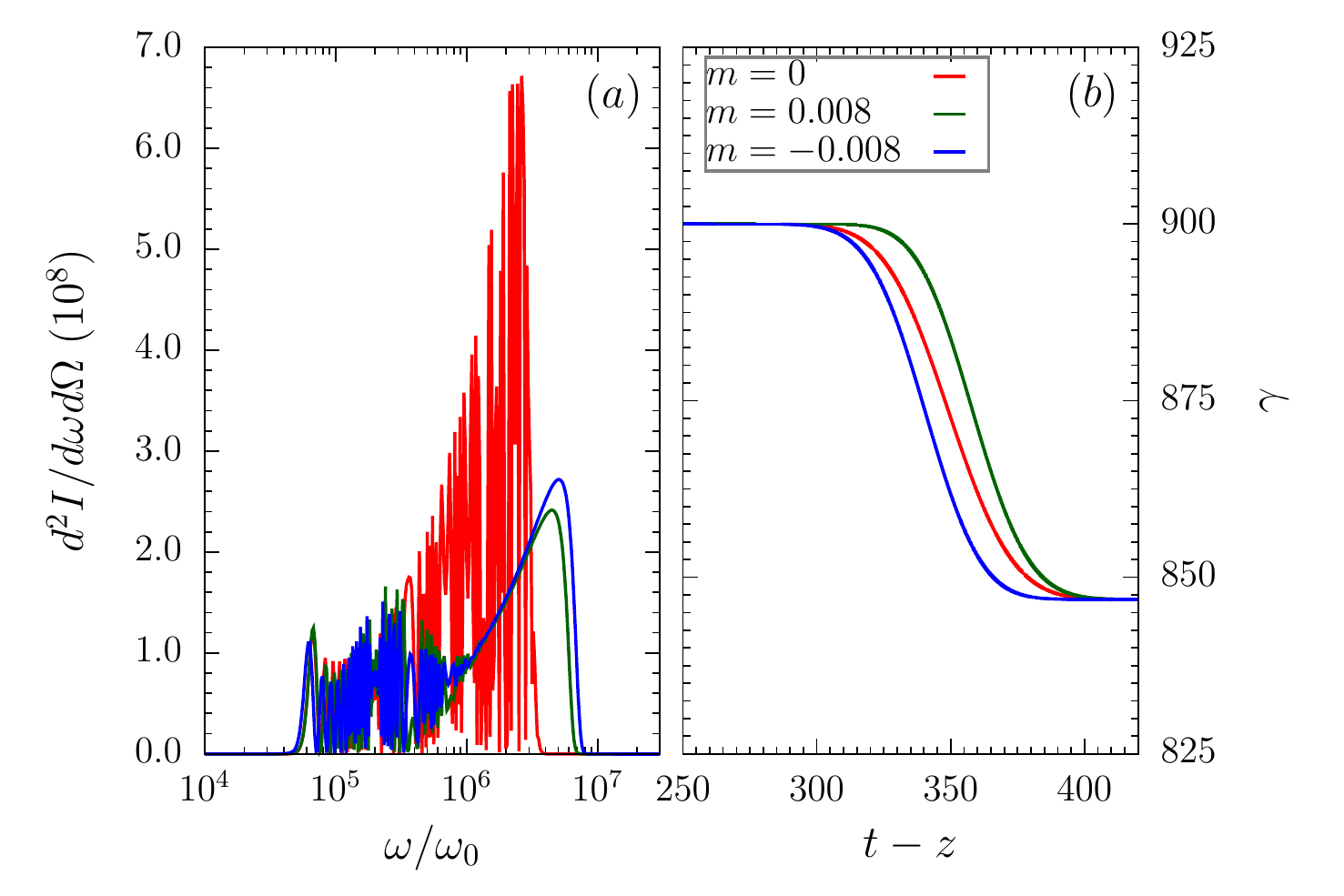}
\caption{Plots showing the time evolution of the electron $\gamma$-factor in chirped and unchirped pulses (chirping constant $b=\pm0.008$). The electron has an initial $\gamma_0=900$ and the laser is a circularly polarised pulse of peak intensity $a_0=10$ and 10 cycles duration FWHM.}  \label{fig:spec_a10} 
\end{figure}

Once we have calculated the particle trajectory in the pulse, the resulting radiation emissions can be obtained via a well-known classical formula.
The energy radiated per unit solid angle per unit frequency is given by \cite{jackson},
 \be 
\frac{d^2I}{d\omega\ d\Omega} 
 = \left|\int\limits_{-\infty}^{\infty}\frac{\mb{n}\times[(\mb{n}-{\beta})\times\dot{\mb{\beta}}]}{(1 - \beta\cdot \mb{n})^2}       
 e^{i\ s [t + D(t)]} dt \right|^2, 
\label{spectrum}
\ee
where $\mb{n}$ is a unit vector pointing from the particle's position to the detector ($D$) located far away from the interaction, and $\beta$ and $\dot{\beta}$ are, respectively, the particle's relativistic velocity  and acceleration. In our dimensionless units, $s = \omega/\omega_0$ is taken to be the harmonic of fundamental frequency. Here  we have normalized the intensity by the factor $e^2/(4\pi^2 c)$. All the quantities in the above equations are evaluated at the retarded time so one can directly do the integration in some finite limit.

\section{Results}
Since this study is concerned with high field intensities, it is instructive to first provide an estimate for when RR effects become important.  Using just the Lorentz force to determine the motion, the radiated power $P$ is given by Larmor's formula in terms of the acceleration,
\begin{eqnarray}
P=\frac{2}{3}\frac{m r_e \textrm{acc}^2}{c}=\frac{2}{3} r_e mc\omega_0^2 a_0^2 \gamma (1+\beta),
\end{eqnarray}
Normalizing this by $\omega m c^2$ we obtain the energy loss per cycle in terms of the electron rest energy $mc^2$ \cite{Koga:2005, Harvey:2011dp}
\begin{eqnarray}
R\equiv \frac{P}{\omega_0 mc^2}=\frac{2}{3}r_e\frac{\omega_0}{c} a_0^2\gamma (1+\beta).\label{R}
\end{eqnarray}
When this parameter reaches unity we are in the ``radiation- dominated regime'' \cite{DiPiazza:2009RR}, where the radiation damping effects are of the same magnitude as the Lorentz force.    

%
We will begin by studying the effects of chirping on the Thomson spectra of a high-energy electron in a moderately intense laser, where RR effects are not very important.  To be specific, we will consider an electron with initial $\gamma_0 = 900$ ($\sim$ 460 MeV) brought into collision with a  circularly polarized laser pulse having peak amplitude $a_0 = 10$ (which corresponds to $2.11\times 10^{20}$ W cm$^{-2}$ for an optical laser). The duration of the unchirped ($b = 0$) laser pulse is taken to be 10 cycles and is reduced when the chirp is introduced in order to maintain a constant energy.  For these parameters we find R=0.0027, meaning that we are in a regime where RR effects are likely to be minimal.
We first solve the Landau Lifshitz equation for these parameters and consider how the $\gamma$-factor changes as the electron passes through the laser pulse.  
In Fig.~\ref{fig:spec_a10}(b) we compare the evolution of the $\gamma$-factor of an unchirped pulse with those of a positive and negative chirp of $b = \pm 0.008$.  It can be seen that the electron in the negatively chirped field loses energy more quickly than that in the unchirped field, and the electron in the positively chirped field less quickly. (However, the overall energy loss is not large, amounting to about 5\% of the starting value.)
In the left hand panel, Fig.~\ref{fig:spec_a10}(a), we show the emission spectra for the three cases.  This has been calculated by inserting the particle trajectories into (\ref{spectrum}).  It can be seen that the electrons in the two chirped pulses emit radiation of a higher energy, but lower amplitude than the electron in the unchirped pulse.  This is not surprising since these electrons pass through regions of higher frequency fields than exist in the unchirped pulse.  We will discuss this in more detail shortly.  Note that the spectra for the two chirped cases are roughly similar.
Observe also that the emission spectra are much cleaner for the electrons in the chirped pulses than in the unchirped pulse.  This is due to the changing frequency causing an overlap of the contributing harmonics and is discussed in Ref.~\cite{Ghebregziabher:2013}.

%
\begin{figure}[t]
\includegraphics[width=1.0\columnwidth]{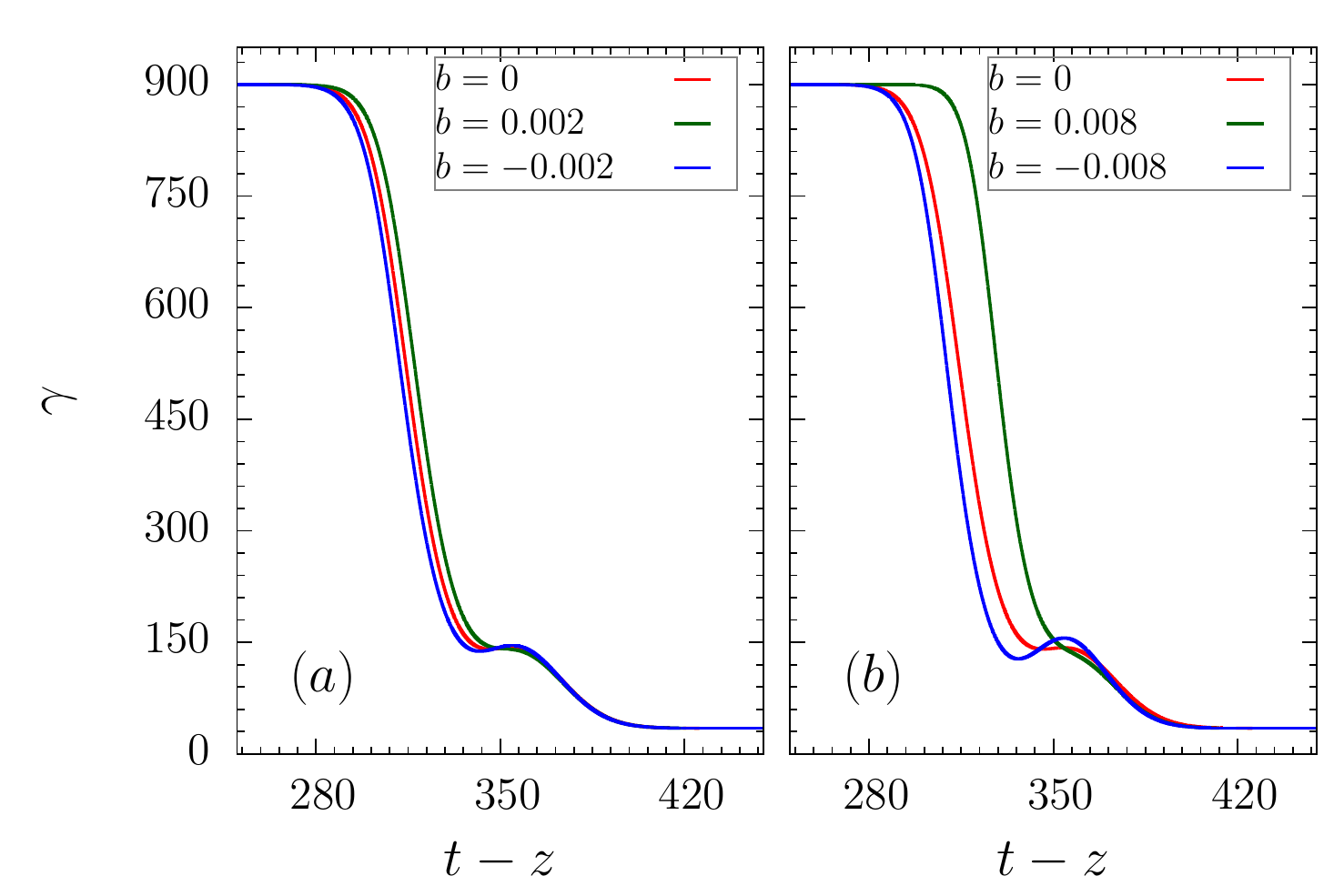}
\caption{Plots showing the time evolution of the electron $\gamma$-factor for two different chirping constants. The electron has an initial $\gamma_0=900$ and the laser is a circularly polarised pulse of peak intensity $a_0=200$ and 10 cycles duration FWHM.  \label{fig:all_gama} }
\end{figure}
Having examined this preliminary example, we now move on to consider a case where RR effects do become important.
For the rest of this study we will work with a circularly polarized laser pulse having peak amplitude $a_0 = 200$ (which corresponds to $8.56\times 10^{22}$ W cm$^{-2}$ for an optical laser), which is just slightly beyond the current state of the art. The counter propagating electron will remain at $\gamma_0 = 900$ ($\sim$ 460 MeV).
For these parameters $R=1.06$, placing us within the regime where RR effects can be expected to dominate.

\begin{figure}[t]
\includegraphics[width=1\columnwidth]{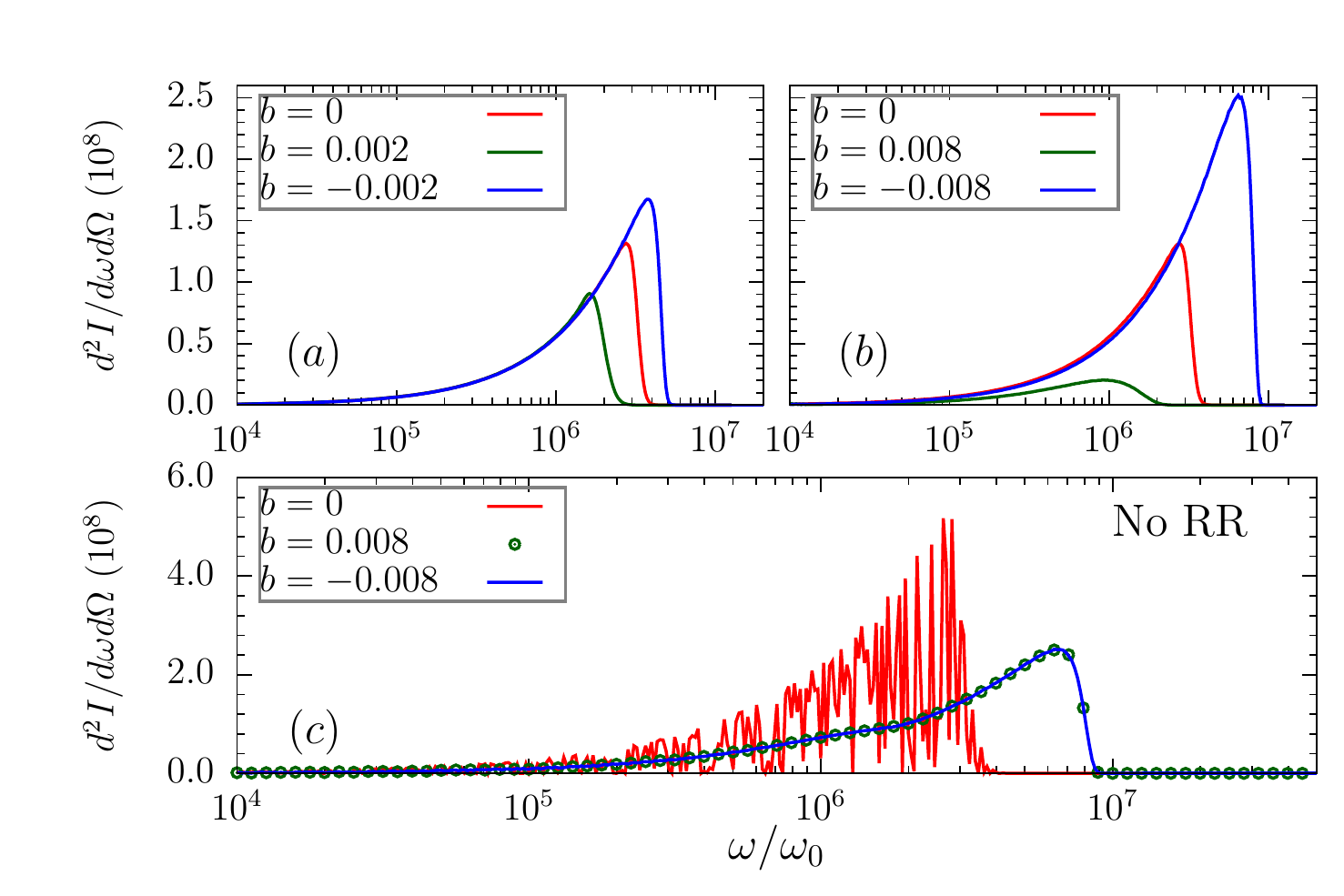}
\caption{Plots showing the emission spectra, evaluated in the backscattering direction ($\theta=180^\circ$) for the two chirp factors (a) and (b). The emission spectra without RR taken into account is also presented (c). Parameters as for Fig.~\ref{fig:all_gama} \label{fig:spectra}. }
\end{figure}
We consider once again how the $\gamma$-factor changes as the electron passes through the laser pulse.  
In Fig.~\ref{fig:all_gama} we show the evolution of $\gamma$ for two different chirping constants,  $b = \pm 0.002$ and $b = \pm 0.008$.  It can be observed, just as in the previous example, that the energy of an electron colliding with a negatively chirped laser pulse (i.e. one where the high frequency components hit the particle first) falls more rapidly by virtue of RR as compared to both its positively (low frequency components first) or unchirped pulse counterparts.
We note that the simple analysis used in eq.~(\ref{R}) is not sufficient to explain this behaviour.  There we were implicitly assuming that the chirped frequency $\omega$ and $\gamma$-factor can be approximated as constant over a laser cycle, which gives $R\sim r_e \gamma e^2 \vert \mathbf{E}\vert^2/\omega m^2c^3$ for $\gamma\gg1$.  Thus, according to this simple model, increasing the pulse frequency will result in a smaller energy loss {\it per cycle}.  However, the electron will pass through a correspondingly larger number of cycles during a given time period and thus the total energy loss over this time period will be approximately the same.  Of course, this is only a crude estimate and we find that, when we calculate the particle motion numerically, the higher the frequency components at the front of the pulse, the faster the particle loses energy.

\begin{figure}[b]
\includegraphics[width=1\columnwidth]{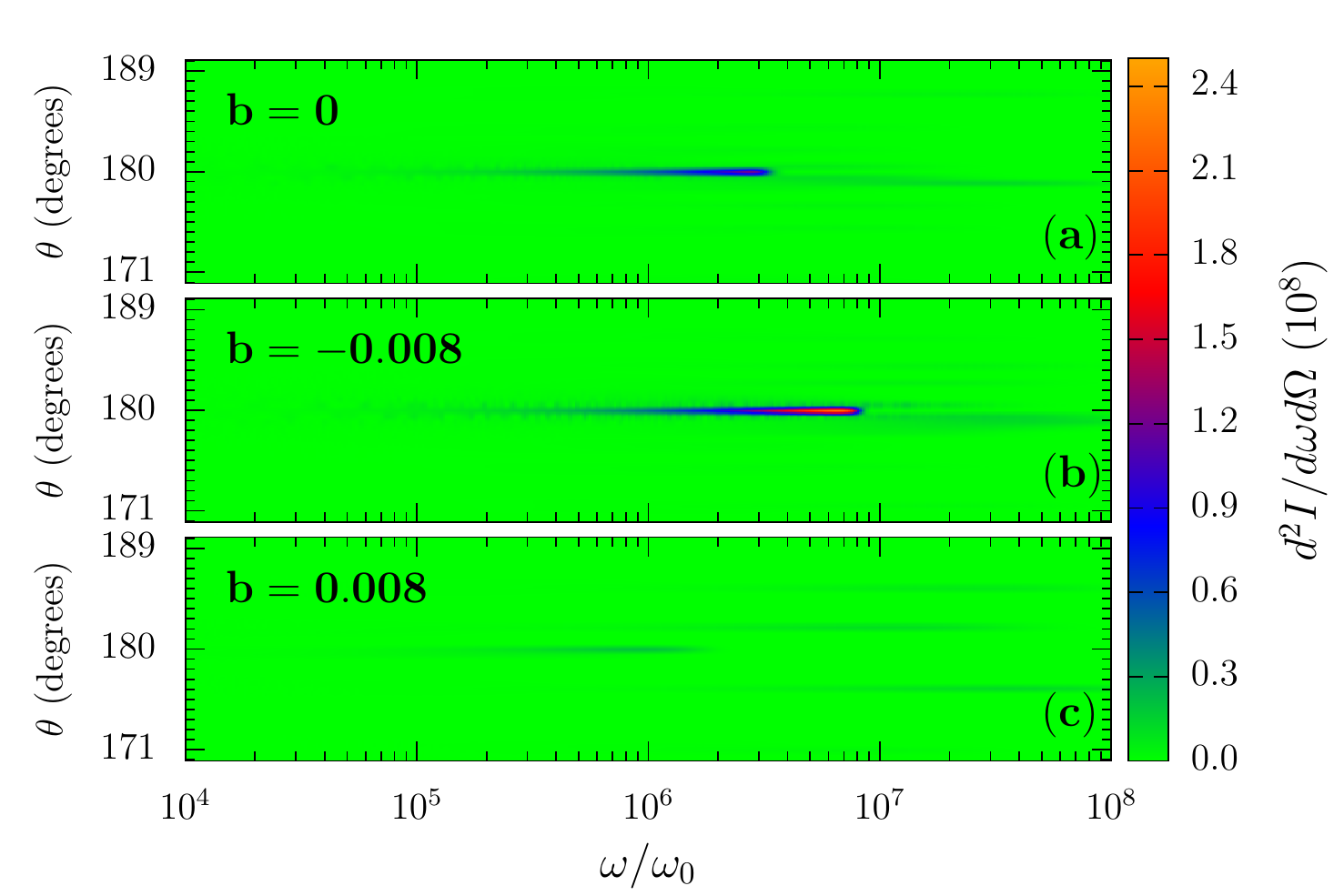}
\caption{Angular distribution of the radiation spectra for $b = 0$ (a), $b = -0.008$ (b) and $b= 0.008$ (c). Parameters as for Fig.~\ref{fig:all_gama}.}
\label{fig:spec3d}
\end{figure}

In Fig.~\ref{fig:spectra} we consider the emission spectra, calculated by inserting the particle trajectories into (\ref{spectrum}).  At high intensities the total emission spectrum will be comprised of a sum of harmonics, corresponding to multiples of the laser frequency. In the unchirped case of constant frequency the properties of the spectrum are well understood. From conservation of momentum arguments it can be shown that the frequency $\omega^\prime_n$ of the $n{th}$ harmonic in the backscattering direction ($\theta=180^\circ$) is given by \cite{PhysRevA.79.063407}
\begin{eqnarray}
\omega^\prime_n\approx\frac{4\gamma^2n\omega_0}{1+a_0^2}. 
\label{omegaprime}
\end{eqnarray}
For large $a_0$ the spectrum begins to decay after the critical harmonic with number $n_\textrm{crit}\sim 3a_0^3/2$ \cite{Esarey:1993}, which for our parameters is $n_\textrm{crit}\sim 10^7$.

\begin{figure}[t]
\includegraphics[width=1\columnwidth]{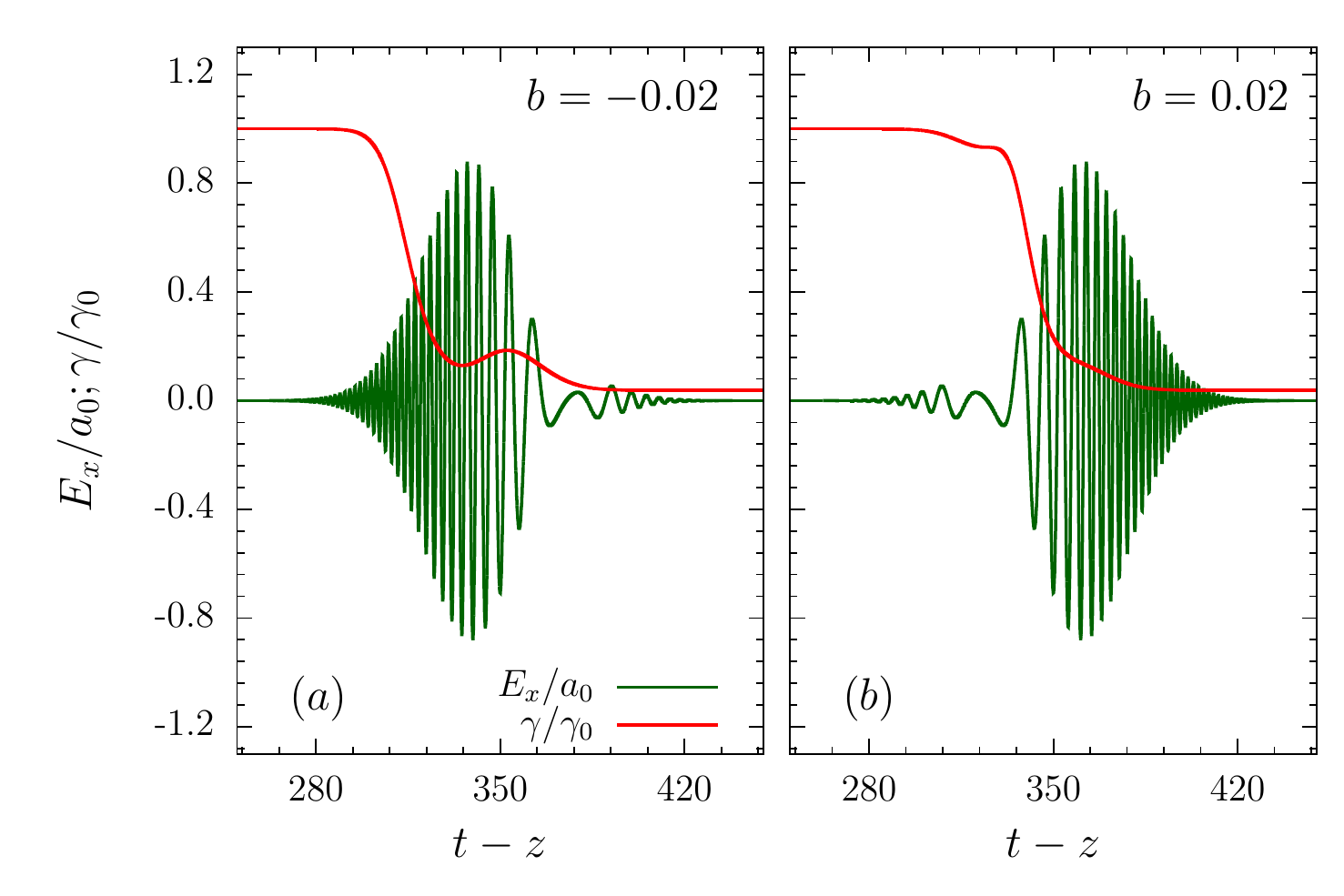}
\caption{Evolution of $\gamma$ along with laser pulse profile for $b = -0.02$ (a) and $b = 0.02$ (b) is presented.  Note that the chirp is unrealistically large in this case and has been chosen to aid illustration. Other parameters the same as for Fig.~\ref{fig:all_gama}. }
\label{fig:high_chirp}
\end{figure}

In our case the situation is more complicated than that of a monochromatic plane wave, but nevertheless we are able to make some very rough analogies to help us understand the physics. 
(For a detailed discussion of how the monochromatic spectra relate to those of a pulsed field we refer the reader to Refs.~\cite{Seipt:2012, Boca:2009, Harvey:2012ie}.)
We can reasonably assume that the spectrum will be dominated by the emissions from the region just to the front of the pulse peak, where both the amplitude and the $\gamma$-factor are relatively large.  
Assuming, for the purposes of this heuristic argument, that the field can be approximated over a given cycle by a monochromatic field with a the same amplitude, we can consider the above expressions in terms of the {\it local} dimensionless intensity $a= e\vert \mathbf{E}\vert/\omega mc$.  In particular, we see that the number of harmonics comprising the spectrum $n_\textrm{crit}$ will then be dependent on the local frequency.
Thus, in the case of the positive chirp, although the frequencies of each harmonic comprising the spectrum will be blue shifted by the larger $\gamma$-factor, the total number of harmonics will be decreased due to the lower frequency, resulting in a total spectrum that covers a smaller energy range. 
Similarly, in the case of the negative chirp, the lower $\gamma$-factor will red-shift the harmonic frequencies, but the total spectrum will be composed of a larger number of harmonics due to the higher frequency of the field.  Thus the total spectrum will span a wider range of energies, as can be seen in the plots.
Finally, the faster rate of energy loss for the negatively chirped case results in a stronger peak signal in the spectrum.

The above argument can be supported by returning to our initial example where RR effects were insignificant.   
In this case no matter whether the chirp was positive or negative, the electron passed through the high frequency portion of the field with an energy comparable to its initial energy, without having lost much energy to RR. Thus it did not matter if it saw the high frequency part of the pulse first or later.   This is why the radiation spectra was more or less identical for both cases (Fig.~\ref{fig:spec_a10}). 
Additionally, we can also consider the current example without the effects of RR (i.e.~by solving the Lorentz equation instead of the Landau Lifshitz equation to determine the particle dynamics).
The resulting spectra are plotted in Fig.~\ref{fig:spectra}(c) where it can be seen that the positive and negative chirp parameters produce identical spectra.

Finally, a more complete presentation of the emission spectra is given in Fig.~\ref{fig:spec3d}, for the chirping constant $b = \pm 0.008$. It can be seen from this figure that the spectra mainly consist of back scattered radiation, i.e. along $\theta = 180^\circ$. 

\begin{figure}[t]
\includegraphics*[width=1\columnwidth]{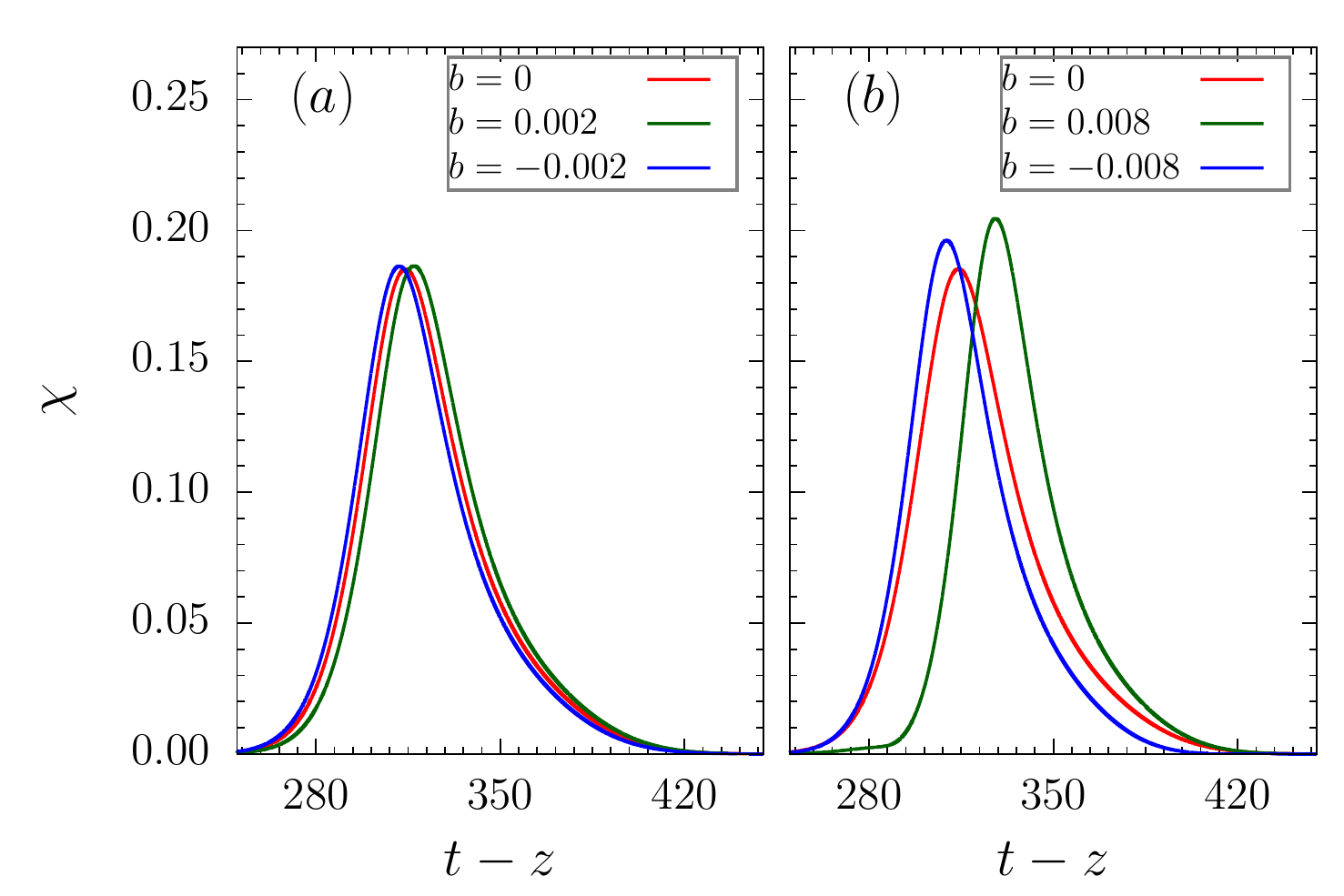}
\caption{Plots showing the time evolution of the quantum efficiency parameter, $\chi$, for two different chirping constants.  Parameters are the same as for Fig.~\ref{fig:all_gama}.}
\label{fig:chiplot}
\end{figure}

In order to illustrate the the effect of the chirp more clearly, the interaction involving a slightly higher chirping constant ($b = \pm 0.02$) is presented in Fig.~\ref{fig:high_chirp}.  In panels (a) and (b) we show how the $\gamma$-factor of the electron evolves as it passes through the laser field.  It can be seen that in the case of negative chirp (panel (a)) the electron has a much higher $\gamma$-factor in the high frequency part of the field than it does for the case of positive chirp (panel (b)).  This results in a higher energy emission spectrum for the negative chirp compared to the positive chirp.  (Note that the results presented in Fig.~\ref{fig:high_chirp} are only for the purpose of demonstrating the chirp.  The energy loss in the early stages of the interaction with the positively chirped pulse is due to the `kink' in the tail of the field.  This is an unphysical artefact caused by our chirp being too large in this illustrative example.)

Since the intensity we have been considering is fairly high, it is worthwhile investigating whether it is indeed valid to treat the system classically, or whether quantum effects should be taken into consideration.  We can measure the importance of quantum effects by considering the dimensionless `quantum efficiency parameter' \cite{Ritus:1985}, which with our normalizations can be written
$\chi=\hbar\omega_0\gamma\sqrt{(\mb{E}+\mb{v}\times\mb{B})^2}/mc^2\sim \gamma E/E_S$, 
where $E_S=1.3\times10^{16}$V/cm is the QED `critical' field (`Sauter-Schwinger' field \cite{QEDcriticalfield1,QEDcriticalfield2,QEDcriticalfield3}).  This parameter describes the work done by the laser field on the particle over of a Compton wavelength. When $\chi\sim 1$ quantum effects will start to dominate, with processes such as vacuum pair production occurring. 
Not only is the calculation of this parameter important in determining the validity of our modelling, it is also of interest to see if the introduction of a chirp can significantly alter its value.
We have already seen how the negative chirp results in an increase in radiated energy, and so it is worth investigating whether the same mechanism will increase the importance of QED effects.  In Fig.~\ref{fig:chiplot} we plot the evolution of the $\chi$ parameter for the cases we have been considering.  It can be seen that $\chi$ reaches a peak value of roughly 0.2 meaning that we are on the threshold of where QED effects are likely to be detectable, but not significant \cite{Harvey:2015rca}.  Thus we are justified in performing our analysis classically. We also see that, while the negative chirp does result in a slight increase in $\chi$, the effect of chirping on QED effects is fairly minimal.  This means that chirping is unlikely to prove a useful tool for probing intensity effects in strong field QED.

\section{Concluding Remarks}

In this article we have investigated the dynamics, and resulting Thomson spectra, of an electron in an intense chirped laser pulse.  Because of RR effects the electron loses energy quickly upon entering the pulse.  This means that by the time the electron reaches the most intense part of the field it has a much lower energy than it began with.  The result is a reduction in energy and brilliance of the emitted Thomson radiation.  By introducing a small, negative chirp into the laser pulse we have shown that it is possible to have the electron enter the region of the pulse where the emissions will be strongest while it still has a large proportion of its initial energy.  For the modest chirp parameters that we have considered this can result in a more than doubling of both the maximum frequency and amplitude of the radiation spectrum a compared to the case of an unchirped field.  This is of great importance in the context of Thomson/Compton scattering experiments using the next generation of ultra-intense laser sources. 
 
\section*{Acknowledgements}
AH acknowledges the Science and Engineering Research Board, Department of Science and Technology, Government of India for funding the project SR/FTP/PS-189/2012.  CH and MM acknowledge support from the Swedish Research Council, grants 2012-5644 and 2013-4248.  The authors thank Olle Lundh for useful discussions.


\begin{thebibliography}{46}%
\makeatletter
\providecommand \@ifxundefined [1]{%
 \@ifx{#1\undefined}
}%
\providecommand \@ifnum [1]{%
 \ifnum #1\expandafter \@firstoftwo
 \else \expandafter \@secondoftwo
 \fi
}%
\providecommand \@ifx [1]{%
 \ifx #1\expandafter \@firstoftwo
 \else \expandafter \@secondoftwo
 \fi
}%
\providecommand \natexlab [1]{#1}%
\providecommand \enquote  [1]{``#1''}%
\providecommand \bibnamefont  [1]{#1}%
\providecommand \bibfnamefont [1]{#1}%
\providecommand \citenamefont [1]{#1}%
\providecommand \href@noop [0]{\@secondoftwo}%
\providecommand \href [0]{\begingroup \@sanitize@url \@href}%
\providecommand \@href[1]{\@@startlink{#1}\@@href}%
\providecommand \@@href[1]{\endgroup#1\@@endlink}%
\providecommand \@sanitize@url [0]{\catcode `\\12\catcode `\$12\catcode
  `\&12\catcode `\#12\catcode `\^12\catcode `\_12\catcode `\%12\relax}%
\providecommand \@@startlink[1]{}%
\providecommand \@@endlink[0]{}%
\providecommand \url  [0]{\begingroup\@sanitize@url \@url }%
\providecommand \@url [1]{\endgroup\@href {#1}{\urlprefix }}%
\providecommand \urlprefix  [0]{URL }%
\providecommand \Eprint [0]{\href }%
\providecommand \doibase [0]{http://dx.doi.org/}%
\providecommand \selectlanguage [0]{\@gobble}%
\providecommand \bibinfo  [0]{\@secondoftwo}%
\providecommand \bibfield  [0]{\@secondoftwo}%
\providecommand \translation [1]{[#1]}%
\providecommand \BibitemOpen [0]{}%
\providecommand \bibitemStop [0]{}%
\providecommand \bibitemNoStop [0]{.\EOS\space}%
\providecommand \EOS [0]{\spacefactor3000\relax}%
\providecommand \BibitemShut  [1]{\csname bibitem#1\endcsname}%
\let\auto@bib@innerbib\@empty
\bibitem [{\citenamefont {Strickland}\ and\ \citenamefont
  {Mourou}(1985{\natexlab{a}})}]{Strickland1985219}%
  \BibitemOpen
  \bibfield  {author} {\bibinfo {author} {\bibfnamefont {D.}~\bibnamefont
  {Strickland}}\ and\ \bibinfo {author} {\bibfnamefont {G.}~\bibnamefont
  {Mourou}},\ }\href {\doibase http://dx.doi.org/10.1016/0030-4018(85)90120-8}
  {\bibfield  {journal} {\bibinfo  {journal} {Optics Communications}\ }\textbf
  {\bibinfo {volume} {56}},\ \bibinfo {pages} {219 } (\bibinfo {year}
  {1985}{\natexlab{a}})}\BibitemShut {NoStop}%
\bibitem [{\citenamefont {Tajima}\ and\ \citenamefont
  {Mourou}(2002)}]{PhysRevSTAB.5.031301}%
  \BibitemOpen
  \bibfield  {author} {\bibinfo {author} {\bibfnamefont {T.}~\bibnamefont
  {Tajima}}\ and\ \bibinfo {author} {\bibfnamefont {G.}~\bibnamefont
  {Mourou}},\ }\href {\doibase 10.1103/PhysRevSTAB.5.031301} {\bibfield
  {journal} {\bibinfo  {journal} {Phys. Rev. ST Accel. Beams}\ }\textbf
  {\bibinfo {volume} {5}},\ \bibinfo {pages} {031301} (\bibinfo {year}
  {2002})}\BibitemShut {NoStop}%
\bibitem [{\citenamefont {Yanovsky}\ \emph {et~al.}(2008)\citenamefont
  {Yanovsky}, \citenamefont {Chvykov}, \citenamefont {Kalinchenko},
  \citenamefont {Rousseau}, \citenamefont {Planchon}, \citenamefont {Matsuoka},
  \citenamefont {Maksimchuk}, \citenamefont {Nees}, \citenamefont {Cheriaux},
  \citenamefont {Mourou},\ and\ \citenamefont {Krushelnick}}]{Yanovsky:2008}%
  \BibitemOpen
  \bibfield  {author} {\bibinfo {author} {\bibfnamefont {V.}~\bibnamefont
  {Yanovsky}}, \bibinfo {author} {\bibfnamefont {V.}~\bibnamefont {Chvykov}},
  \bibinfo {author} {\bibfnamefont {G.}~\bibnamefont {Kalinchenko}}, \bibinfo
  {author} {\bibfnamefont {P.}~\bibnamefont {Rousseau}}, \bibinfo {author}
  {\bibfnamefont {T.}~\bibnamefont {Planchon}}, \bibinfo {author}
  {\bibfnamefont {T.}~\bibnamefont {Matsuoka}}, \bibinfo {author}
  {\bibfnamefont {A.}~\bibnamefont {Maksimchuk}}, \bibinfo {author}
  {\bibfnamefont {J.}~\bibnamefont {Nees}}, \bibinfo {author} {\bibfnamefont
  {G.}~\bibnamefont {Cheriaux}}, \bibinfo {author} {\bibfnamefont
  {G.}~\bibnamefont {Mourou}}, \ and\ \bibinfo {author} {\bibfnamefont
  {K.}~\bibnamefont {Krushelnick}},\ }\href {\doibase 10.1364/OE.16.002109}
  {\bibfield  {journal} {\bibinfo  {journal} {Opt. Express}\ }\textbf {\bibinfo
  {volume} {16}},\ \bibinfo {pages} {2109} (\bibinfo {year}
  {2008})}\BibitemShut {NoStop}%
\bibitem [{Vul()}]{Vulcan}%
  \BibitemOpen
  \href {{https://www.stfc.ac.uk/CLF/12248.aspx}} {\enquote {\bibinfo {title}
  {The vulcan 10 petawatt project},}\ }\BibitemShut {NoStop}%
\bibitem [{ELI()}]{ELI}%
  \BibitemOpen
  \href {{http://www.eli-laser.eu/index.html}} {\enquote {\bibinfo {title} {The
  extreme light infrastructure (eli) project},}\ }\BibitemShut {NoStop}%
\bibitem [{XCE()}]{XCELS}%
  \BibitemOpen
  \href {{http://www.xcels.iapras.ru/}} {\enquote {\bibinfo {title} {The
  exawatt center for extreme light studies (xcels)},}\ }\BibitemShut {NoStop}%
\bibitem [{\citenamefont {Wu}()}]{Wu}%
  \BibitemOpen
  \bibfield  {author} {\bibinfo {author} {\bibfnamefont {Y.~K.}\ \bibnamefont
  {Wu}},\ }\href@noop {} {\emph {\bibinfo {title} {Overview of High Intensity
  Gamma-ray Source- Capabilities and Future Upgrades, 2013 International
  Workshop on Polarized Sources, Targets and Polarimetry (University of
  Virginia, 2013).}}}\BibitemShut {Stop}%
\bibitem [{\citenamefont {Lawrence}\ \emph {et~al.}(2008)\citenamefont
  {Lawrence}, \citenamefont {Ten~Haken},\ and\ \citenamefont
  {Giaccia}}]{Lawrence}%
  \BibitemOpen
  \bibfield  {author} {\bibinfo {author} {\bibfnamefont {T.~S.}\ \bibnamefont
  {Lawrence}}, \bibinfo {author} {\bibfnamefont {R.~K.}\ \bibnamefont
  {Ten~Haken}}, \ and\ \bibinfo {author} {\bibfnamefont {A.}~\bibnamefont
  {Giaccia}},\ }in\ \href@noop {} {\emph {\bibinfo {booktitle} {Principles and
  Practice of Oncology}}}\ (\bibinfo  {publisher} {Lippincott, Williams, and
  Wilkins},\ \bibinfo {address} {Philadelphia},\ \bibinfo {year} {2008})\
  \bibinfo {edition} {8th}\ ed.\BibitemShut {Stop}%
\bibitem [{\citenamefont {Glinec}\ \emph {et~al.}(2005)\citenamefont {Glinec},
  \citenamefont {Faure}, \citenamefont {Dain}, \citenamefont {Darbon},
  \citenamefont {Hosokai}, \citenamefont {Santos}, \citenamefont {Lefebvre},
  \citenamefont {Rousseau}, \citenamefont {Burgy}, \citenamefont {Mercier},\
  and\ \citenamefont {Malka}}]{PhysRevLett.94.025003}%
  \BibitemOpen
  \bibfield  {author} {\bibinfo {author} {\bibfnamefont {Y.}~\bibnamefont
  {Glinec}}, \bibinfo {author} {\bibfnamefont {J.}~\bibnamefont {Faure}},
  \bibinfo {author} {\bibfnamefont {L.~L.}\ \bibnamefont {Dain}}, \bibinfo
  {author} {\bibfnamefont {S.}~\bibnamefont {Darbon}}, \bibinfo {author}
  {\bibfnamefont {T.}~\bibnamefont {Hosokai}}, \bibinfo {author} {\bibfnamefont
  {J.~J.}\ \bibnamefont {Santos}}, \bibinfo {author} {\bibfnamefont
  {E.}~\bibnamefont {Lefebvre}}, \bibinfo {author} {\bibfnamefont {J.~P.}\
  \bibnamefont {Rousseau}}, \bibinfo {author} {\bibfnamefont {F.}~\bibnamefont
  {Burgy}}, \bibinfo {author} {\bibfnamefont {B.}~\bibnamefont {Mercier}}, \
  and\ \bibinfo {author} {\bibfnamefont {V.}~\bibnamefont {Malka}},\ }\href
  {\doibase 10.1103/PhysRevLett.94.025003} {\bibfield  {journal} {\bibinfo
  {journal} {Phys. Rev. Lett.}\ }\textbf {\bibinfo {volume} {94}},\ \bibinfo
  {pages} {025003} (\bibinfo {year} {2005})}\BibitemShut {NoStop}%
\bibitem [{\citenamefont {{Cipiccia}}\ \emph {et~al.}(2011)\citenamefont
  {{Cipiccia}}, \citenamefont {{Islam}}, \citenamefont {{Ersfeld}},
  \citenamefont {{Shanks}}, \citenamefont {{Brunetti}}, \citenamefont
  {{Vieux}}, \citenamefont {{Yang}}, \citenamefont {{Issac}}, \citenamefont
  {{Wiggins}}, \citenamefont {{Welsh}}, \citenamefont {{Anania}}, \citenamefont
  {{Maneuski}}, \citenamefont {{Montgomery}}, \citenamefont {{Smith}},
  \citenamefont {{Hoek}}, \citenamefont {{Hamilton}}, \citenamefont {{Lemos}},
  \citenamefont {{Symes}}, \citenamefont {{Rajeev}}, \citenamefont {{Shea}},
  \citenamefont {{Dias}},\ and\ \citenamefont
  {{Jaroszynski}}}]{2011NatPh...7..867C}%
  \BibitemOpen
  \bibfield  {author} {\bibinfo {author} {\bibfnamefont {S.}~\bibnamefont
  {{Cipiccia}}}, \bibinfo {author} {\bibfnamefont {M.~R.}\ \bibnamefont
  {{Islam}}}, \bibinfo {author} {\bibfnamefont {B.}~\bibnamefont {{Ersfeld}}},
  \bibinfo {author} {\bibfnamefont {R.~P.}\ \bibnamefont {{Shanks}}}, \bibinfo
  {author} {\bibfnamefont {E.}~\bibnamefont {{Brunetti}}}, \bibinfo {author}
  {\bibfnamefont {G.}~\bibnamefont {{Vieux}}}, \bibinfo {author} {\bibfnamefont
  {X.}~\bibnamefont {{Yang}}}, \bibinfo {author} {\bibfnamefont {R.~C.}\
  \bibnamefont {{Issac}}}, \bibinfo {author} {\bibfnamefont {S.~M.}\
  \bibnamefont {{Wiggins}}}, \bibinfo {author} {\bibfnamefont {G.~H.}\
  \bibnamefont {{Welsh}}}, \bibinfo {author} {\bibfnamefont {M.-P.}\
  \bibnamefont {{Anania}}}, \bibinfo {author} {\bibfnamefont {D.}~\bibnamefont
  {{Maneuski}}}, \bibinfo {author} {\bibfnamefont {R.}~\bibnamefont
  {{Montgomery}}}, \bibinfo {author} {\bibfnamefont {G.}~\bibnamefont
  {{Smith}}}, \bibinfo {author} {\bibfnamefont {M.}~\bibnamefont {{Hoek}}},
  \bibinfo {author} {\bibfnamefont {D.~J.}\ \bibnamefont {{Hamilton}}},
  \bibinfo {author} {\bibfnamefont {N.~R.~C.}\ \bibnamefont {{Lemos}}},
  \bibinfo {author} {\bibfnamefont {D.}~\bibnamefont {{Symes}}}, \bibinfo
  {author} {\bibfnamefont {P.~P.}\ \bibnamefont {{Rajeev}}}, \bibinfo {author}
  {\bibfnamefont {V.~O.}\ \bibnamefont {{Shea}}}, \bibinfo {author}
  {\bibfnamefont {J.~M.}\ \bibnamefont {{Dias}}}, \ and\ \bibinfo {author}
  {\bibfnamefont {D.~A.}\ \bibnamefont {{Jaroszynski}}},\ }\href {\doibase
  10.1038/nphys2090} {\bibfield  {journal} {\bibinfo  {journal} {Nature
  Physics}\ }\textbf {\bibinfo {volume} {7}},\ \bibinfo {pages} {867} (\bibinfo
  {year} {2011})}\BibitemShut {NoStop}%
\bibitem [{\citenamefont {Sarri}\ \emph {et~al.}(2014)\citenamefont {Sarri},
  \citenamefont {Corvan}, \citenamefont {Schumaker}, \citenamefont {Cole},
  \citenamefont {Di~Piazza}, \citenamefont {Ahmed}, \citenamefont {Harvey},
  \citenamefont {Keitel}, \citenamefont {Krushelnick}, \citenamefont {Mangles},
  \citenamefont {Najmudin}, \citenamefont {Symes}, \citenamefont {Thomas},
  \citenamefont {Yeung}, \citenamefont {Zhao},\ and\ \citenamefont
  {Zepf}}]{PhysRevLett.113.224801}%
  \BibitemOpen
  \bibfield  {author} {\bibinfo {author} {\bibfnamefont {G.}~\bibnamefont
  {Sarri}}, \bibinfo {author} {\bibfnamefont {D.~J.}\ \bibnamefont {Corvan}},
  \bibinfo {author} {\bibfnamefont {W.}~\bibnamefont {Schumaker}}, \bibinfo
  {author} {\bibfnamefont {J.~M.}\ \bibnamefont {Cole}}, \bibinfo {author}
  {\bibfnamefont {A.}~\bibnamefont {Di~Piazza}}, \bibinfo {author}
  {\bibfnamefont {H.}~\bibnamefont {Ahmed}}, \bibinfo {author} {\bibfnamefont
  {C.}~\bibnamefont {Harvey}}, \bibinfo {author} {\bibfnamefont {C.~H.}\
  \bibnamefont {Keitel}}, \bibinfo {author} {\bibfnamefont {K.}~\bibnamefont
  {Krushelnick}}, \bibinfo {author} {\bibfnamefont {S.~P.~D.}\ \bibnamefont
  {Mangles}}, \bibinfo {author} {\bibfnamefont {Z.}~\bibnamefont {Najmudin}},
  \bibinfo {author} {\bibfnamefont {D.}~\bibnamefont {Symes}}, \bibinfo
  {author} {\bibfnamefont {A.~G.~R.}\ \bibnamefont {Thomas}}, \bibinfo {author}
  {\bibfnamefont {M.}~\bibnamefont {Yeung}}, \bibinfo {author} {\bibfnamefont
  {Z.}~\bibnamefont {Zhao}}, \ and\ \bibinfo {author} {\bibfnamefont
  {M.}~\bibnamefont {Zepf}},\ }\href {\doibase 10.1103/PhysRevLett.113.224801}
  {\bibfield  {journal} {\bibinfo  {journal} {Phys. Rev. Lett.}\ }\textbf
  {\bibinfo {volume} {113}},\ \bibinfo {pages} {224801} (\bibinfo {year}
  {2014})}\BibitemShut {NoStop}%
\bibitem [{\citenamefont {Heinzl}(2012)}]{Heinzl:2011ur}%
  \BibitemOpen
  \bibfield  {author} {\bibinfo {author} {\bibfnamefont {T.}~\bibnamefont
  {Heinzl}},\ }\href {\doibase 10.1142/S2010194512007283,
  10.1142/S0217751X1260010X} {\bibfield  {journal} {\bibinfo  {journal}
  {Int.~J.~Mod.~Phys.}\ }\textbf {\bibinfo {volume} {A27}},\ \bibinfo {pages}
  {1260010} (\bibinfo {year} {2012})}\BibitemShut {NoStop}%
\bibitem [{\citenamefont {Di~Piazza}\ \emph {et~al.}(2012)\citenamefont
  {Di~Piazza}, \citenamefont {M\"uller}, \citenamefont {Hatsagortsyan},\ and\
  \citenamefont {Keitel}}]{DiPiazza:2012RevModPhys}%
  \BibitemOpen
  \bibfield  {author} {\bibinfo {author} {\bibfnamefont {A.}~\bibnamefont
  {Di~Piazza}}, \bibinfo {author} {\bibfnamefont {C.}~\bibnamefont {M\"uller}},
  \bibinfo {author} {\bibfnamefont {K.~Z.}\ \bibnamefont {Hatsagortsyan}}, \
  and\ \bibinfo {author} {\bibfnamefont {C.~H.}\ \bibnamefont {Keitel}},\
  }\href {\doibase 10.1103/RevModPhys.84.1177} {\bibfield  {journal} {\bibinfo
  {journal} {Rev. Mod. Phys.}\ }\textbf {\bibinfo {volume} {84}},\ \bibinfo
  {pages} {1177} (\bibinfo {year} {2012})}\BibitemShut {NoStop}%
\bibitem [{\citenamefont {Kravets}\ \emph {et~al.}(2013)\citenamefont
  {Kravets}, \citenamefont {Noble},\ and\ \citenamefont
  {Jaroszynski}}]{Kravets:2013}%
  \BibitemOpen
  \bibfield  {author} {\bibinfo {author} {\bibfnamefont {Y.}~\bibnamefont
  {Kravets}}, \bibinfo {author} {\bibfnamefont {A.}~\bibnamefont {Noble}}, \
  and\ \bibinfo {author} {\bibfnamefont {D.}~\bibnamefont {Jaroszynski}},\
  }\href {\doibase 10.1103/PhysRevE.88.011201} {\bibfield  {journal} {\bibinfo
  {journal} {Phys. Rev. E}\ }\textbf {\bibinfo {volume} {88}},\ \bibinfo
  {pages} {011201} (\bibinfo {year} {2013})}\BibitemShut {NoStop}%
\bibitem [{\citenamefont {Harvey}\ and\ \citenamefont
  {Marklund}(2012)}]{Harvey:2011mp}%
  \BibitemOpen
  \bibfield  {author} {\bibinfo {author} {\bibfnamefont {C.}~\bibnamefont
  {Harvey}}\ and\ \bibinfo {author} {\bibfnamefont {M.}~\bibnamefont
  {Marklund}},\ }\href {\doibase 10.1103/PhysRevA.85.013412} {\bibfield
  {journal} {\bibinfo  {journal} {Phys. Rev. A}\ }\textbf {\bibinfo {volume}
  {85}},\ \bibinfo {pages} {013412} (\bibinfo {year} {2012})}\BibitemShut
  {NoStop}%
\bibitem [{\citenamefont {Heinzl}\ \emph {et~al.}(2015)\citenamefont {Heinzl},
  \citenamefont {Harvey}, \citenamefont {Ilderton}, \citenamefont {Marklund},
  \citenamefont {Bulanov}, \citenamefont {Rykovanov}, \citenamefont
  {Schroeder}, \citenamefont {Esarey},\ and\ \citenamefont
  {Leemans}}]{Heinzl:2015}%
  \BibitemOpen
  \bibfield  {author} {\bibinfo {author} {\bibfnamefont {T.}~\bibnamefont
  {Heinzl}}, \bibinfo {author} {\bibfnamefont {C.}~\bibnamefont {Harvey}},
  \bibinfo {author} {\bibfnamefont {A.}~\bibnamefont {Ilderton}}, \bibinfo
  {author} {\bibfnamefont {M.}~\bibnamefont {Marklund}}, \bibinfo {author}
  {\bibfnamefont {S.~S.}\ \bibnamefont {Bulanov}}, \bibinfo {author}
  {\bibfnamefont {S.}~\bibnamefont {Rykovanov}}, \bibinfo {author}
  {\bibfnamefont {C.~B.}\ \bibnamefont {Schroeder}}, \bibinfo {author}
  {\bibfnamefont {E.}~\bibnamefont {Esarey}}, \ and\ \bibinfo {author}
  {\bibfnamefont {W.~P.}\ \bibnamefont {Leemans}},\ }\href {\doibase
  10.1103/PhysRevE.91.023207} {\bibfield  {journal} {\bibinfo  {journal} {Phys.
  Rev. E}\ }\textbf {\bibinfo {volume} {91}},\ \bibinfo {pages} {023207}
  (\bibinfo {year} {2015})}\BibitemShut {NoStop}%
\bibitem [{\citenamefont {Holkundkar}\ and\ \citenamefont
  {Harvey}(2014)}]{Holkundkar:2014}%
  \BibitemOpen
  \bibfield  {author} {\bibinfo {author} {\bibfnamefont {A.~R.}\ \bibnamefont
  {Holkundkar}}\ and\ \bibinfo {author} {\bibfnamefont {C.}~\bibnamefont
  {Harvey}},\ }\href {\doibase http://dx.doi.org/10.1063/1.4897307} {\bibfield
  {journal} {\bibinfo  {journal} {Physics of Plasmas}\ }\textbf {\bibinfo
  {volume} {21}},\ \bibinfo {eid} {103102} (\bibinfo {year}
  {2014})}\BibitemShut {NoStop}%
\bibitem [{\citenamefont {Hartemann}\ and\ \citenamefont
  {Kerman}(1996)}]{Hartemann:1996}%
  \BibitemOpen
  \bibfield  {author} {\bibinfo {author} {\bibfnamefont {F.~V.}\ \bibnamefont
  {Hartemann}}\ and\ \bibinfo {author} {\bibfnamefont {A.~K.}\ \bibnamefont
  {Kerman}},\ }\href {\doibase 10.1103/PhysRevLett.76.624} {\bibfield
  {journal} {\bibinfo  {journal} {Phys. Rev. Lett.}\ }\textbf {\bibinfo
  {volume} {76}},\ \bibinfo {pages} {624} (\bibinfo {year} {1996})}\BibitemShut
  {NoStop}%
\bibitem [{\citenamefont {Koga}\ \emph {et~al.}(2005)\citenamefont {Koga},
  \citenamefont {Esirkepov},\ and\ \citenamefont {Bulanov}}]{Koga:2005}%
  \BibitemOpen
  \bibfield  {author} {\bibinfo {author} {\bibfnamefont {J.}~\bibnamefont
  {Koga}}, \bibinfo {author} {\bibfnamefont {T.~Z.}\ \bibnamefont {Esirkepov}},
  \ and\ \bibinfo {author} {\bibfnamefont {S.~V.}\ \bibnamefont {Bulanov}},\
  }\href {\doibase http://dx.doi.org/10.1063/1.2013067} {\bibfield  {journal}
  {\bibinfo  {journal} {Physics of Plasmas}\ }\textbf {\bibinfo {volume}
  {12}},\ \bibinfo {eid} {093106} (\bibinfo {year} {2005})}\BibitemShut
  {NoStop}%
\bibitem [{\citenamefont {Yazdani}\ \emph {et~al.}(2014)\citenamefont
  {Yazdani}, \citenamefont {Sadighi-Bonabi}, \citenamefont {Afarideh},
  \citenamefont {Riazi},\ and\ \citenamefont {Hora}}]{Yazdani:2014}%
  \BibitemOpen
  \bibfield  {author} {\bibinfo {author} {\bibfnamefont {E.}~\bibnamefont
  {Yazdani}}, \bibinfo {author} {\bibfnamefont {R.}~\bibnamefont
  {Sadighi-Bonabi}}, \bibinfo {author} {\bibfnamefont {H.}~\bibnamefont
  {Afarideh}}, \bibinfo {author} {\bibfnamefont {Z.}~\bibnamefont {Riazi}}, \
  and\ \bibinfo {author} {\bibfnamefont {H.}~\bibnamefont {Hora}},\ }\href
  {\doibase http://dx.doi.org/10.1063/1.4894777} {\bibfield  {journal}
  {\bibinfo  {journal} {Journal of Applied Physics}\ }\textbf {\bibinfo
  {volume} {116}},\ \bibinfo {eid} {103302} (\bibinfo {year}
  {2014})}\BibitemShut {NoStop}%
\bibitem [{\citenamefont {{Rykovanov}}\ \emph {et~al.}(2014)\citenamefont
  {{Rykovanov}}, \citenamefont {{Geddes}}, \citenamefont {{Schroeder}},
  \citenamefont {{Esarey}},\ and\ \citenamefont {{Leemans}}}]{Rykovanov:2014}%
  \BibitemOpen
  \bibfield  {author} {\bibinfo {author} {\bibfnamefont {S.~G.}\ \bibnamefont
  {{Rykovanov}}}, \bibinfo {author} {\bibfnamefont {C.~G.~R.}\ \bibnamefont
  {{Geddes}}}, \bibinfo {author} {\bibfnamefont {C.~B.}\ \bibnamefont
  {{Schroeder}}}, \bibinfo {author} {\bibfnamefont {E.}~\bibnamefont
  {{Esarey}}}, \ and\ \bibinfo {author} {\bibfnamefont {W.~P.}\ \bibnamefont
  {{Leemans}}},\ }\href@noop {} {\bibfield  {journal} {\bibinfo  {journal}
  {ArXiv e-prints}\ } (\bibinfo {year} {2014})},\ \Eprint
  {http://arxiv.org/abs/1412.2517} {arXiv:1412.2517 [physics.plasm-ph]}
  \BibitemShut {NoStop}%
\bibitem [{\citenamefont {Ghebregziabher}\ \emph {et~al.}(2013)\citenamefont
  {Ghebregziabher}, \citenamefont {Shadwick},\ and\ \citenamefont
  {Umstadter}}]{Ghebregziabher:2013}%
  \BibitemOpen
  \bibfield  {author} {\bibinfo {author} {\bibfnamefont {I.}~\bibnamefont
  {Ghebregziabher}}, \bibinfo {author} {\bibfnamefont {B.~A.}\ \bibnamefont
  {Shadwick}}, \ and\ \bibinfo {author} {\bibfnamefont {D.}~\bibnamefont
  {Umstadter}},\ }\href {\doibase 10.1103/PhysRevSTAB.16.030705} {\bibfield
  {journal} {\bibinfo  {journal} {Phys. Rev. ST Accel. Beams}\ }\textbf
  {\bibinfo {volume} {16}},\ \bibinfo {pages} {030705} (\bibinfo {year}
  {2013})}\BibitemShut {NoStop}%
\bibitem [{\citenamefont {Strickland}\ and\ \citenamefont
  {Mourou}(1985{\natexlab{b}})}]{Strickland1985447}%
  \BibitemOpen
  \bibfield  {author} {\bibinfo {author} {\bibfnamefont {D.}~\bibnamefont
  {Strickland}}\ and\ \bibinfo {author} {\bibfnamefont {G.}~\bibnamefont
  {Mourou}},\ }\href {\doibase http://dx.doi.org/10.1016/0030-4018(85)90151-8}
  {\bibfield  {journal} {\bibinfo  {journal} {Optics Communications}\ }\textbf
  {\bibinfo {volume} {55}},\ \bibinfo {pages} {447 } (\bibinfo {year}
  {1985}{\natexlab{b}})}\BibitemShut {NoStop}%
\bibitem [{\citenamefont {Kamada}\ \emph {et~al.}(2014)\citenamefont {Kamada},
  \citenamefont {Yoshida},\ and\ \citenamefont {Aoki}}]{kamada_chirp}%
  \BibitemOpen
  \bibfield  {author} {\bibinfo {author} {\bibfnamefont {S.}~\bibnamefont
  {Kamada}}, \bibinfo {author} {\bibfnamefont {T.}~\bibnamefont {Yoshida}}, \
  and\ \bibinfo {author} {\bibfnamefont {T.}~\bibnamefont {Aoki}},\ }\href
  {\doibase http://dx.doi.org/10.1063/1.4867983} {\bibfield  {journal}
  {\bibinfo  {journal} {Applied Physics Letters}\ }\textbf {\bibinfo {volume}
  {104}},\ \bibinfo {eid} {101102} (\bibinfo {year} {2014})}\BibitemShut
  {NoStop}%
\bibitem [{\citenamefont {Lorentz}()}]{Lorentz:1905}%
  \BibitemOpen
  \bibfield  {author} {\bibinfo {author} {\bibfnamefont {H.~A.}\ \bibnamefont
  {Lorentz}},\ }\href@noop {} {\emph {\bibinfo {title} {The Theory of
  Electrons}}}\ (\bibinfo  {publisher} {Teubner, Leipzig (1905), reprinted by
  Dover Publications, New York, (1952) and Cosimo, New York,
  (2007)})\BibitemShut {NoStop}%
\bibitem [{\citenamefont {Abraham}(1905)}]{Abraham:1905}%
  \BibitemOpen
  \bibfield  {author} {\bibinfo {author} {\bibfnamefont {M.}~\bibnamefont
  {Abraham}},\ }\href@noop {} {\emph {\bibinfo {title} {Theorie der
  Elektrizit\"at}}}\ (\bibinfo  {publisher} {Teubner, Leipzig},\ \bibinfo
  {year} {1905})\BibitemShut {NoStop}%
\bibitem [{\citenamefont {Dirac}(1938)}]{Dirac:1938nz}%
  \BibitemOpen
  \bibfield  {author} {\bibinfo {author} {\bibfnamefont {P.~A.~M.}\
  \bibnamefont {Dirac}},\ }\href {\doibase 10.1098/rspa.1938.0124} {\bibfield
  {journal} {\bibinfo  {journal} {PRSLA}\ }\textbf {\bibinfo {volume} {167}},\
  \bibinfo {pages} {148} (\bibinfo {year} {1938})}\BibitemShut {NoStop}%
\bibitem [{\citenamefont {Landau}\ and\ \citenamefont {Lifshitz}(1987)}]{LLII}%
  \BibitemOpen
  \bibfield  {author} {\bibinfo {author} {\bibfnamefont {L.~D.}\ \bibnamefont
  {Landau}}\ and\ \bibinfo {author} {\bibfnamefont {E.~M.}\ \bibnamefont
  {Lifshitz}},\ }\href@noop {} {\emph {\bibinfo {title} {The Classical Theory
  of Fields, Course of Theoretical Physics Vol.~2}}}\ (\bibinfo  {publisher}
  {Butterworth-Heinemann, Oxford},\ \bibinfo {year} {1987})\BibitemShut
  {NoStop}%
\bibitem [{\citenamefont {Burton}\ and\ \citenamefont
  {Noble}(2014)}]{Burton:2014}%
  \BibitemOpen
  \bibfield  {author} {\bibinfo {author} {\bibfnamefont {D.~A.}\ \bibnamefont
  {Burton}}\ and\ \bibinfo {author} {\bibfnamefont {A.}~\bibnamefont {Noble}},\
  }\href {\doibase 10.1080/00107514.2014.886840} {\bibfield  {journal}
  {\bibinfo  {journal} {Contemporary Physics}\ }\textbf {\bibinfo {volume}
  {55}},\ \bibinfo {pages} {110} (\bibinfo {year} {2014})}\BibitemShut
  {NoStop}%
\bibitem [{\citenamefont {Vranic}\ \emph {et~al.}(2015)\citenamefont {Vranic},
  \citenamefont {Martins}, \citenamefont {Fonseca},\ and\ \citenamefont
  {Silva}}]{Vranic:2015}%
  \BibitemOpen
  \bibfield  {author} {\bibinfo {author} {\bibfnamefont {M.}~\bibnamefont
  {Vranic}}, \bibinfo {author} {\bibfnamefont {J.~L.}\ \bibnamefont {Martins}},
  \bibinfo {author} {\bibfnamefont {R.~A.}\ \bibnamefont {Fonseca}}, \ and\
  \bibinfo {author} {\bibfnamefont {L.~O.}\ \bibnamefont {Silva}},\ }\href@noop
  {} {\bibfield  {journal} {\bibinfo  {journal} {arXiv preprint
  arXiv:1502.02432}\ } (\bibinfo {year} {2015})}\BibitemShut {NoStop}%
\bibitem [{\citenamefont {Krivitski\v{\i}}\ and\ \citenamefont
  {Tsytovich}(1991)}]{0038-5670-34-3-A04}%
  \BibitemOpen
  \bibfield  {author} {\bibinfo {author} {\bibfnamefont {V.~S.}\ \bibnamefont
  {Krivitski\v{\i}}}\ and\ \bibinfo {author} {\bibfnamefont {V.~N.}\
  \bibnamefont {Tsytovich}},\ }\href
  {http://stacks.iop.org/0038-5670/34/i=3/a=A04} {\bibfield  {journal}
  {\bibinfo  {journal} {Soviet Physics Uspekhi}\ }\textbf {\bibinfo {volume}
  {34}},\ \bibinfo {pages} {250} (\bibinfo {year} {1991})}\BibitemShut
  {NoStop}%
\bibitem [{\citenamefont {Ilderton}\ and\ \citenamefont
  {Torgrimsson}(2013)}]{Ilderton:2013tb}%
  \BibitemOpen
  \bibfield  {author} {\bibinfo {author} {\bibfnamefont {A.}~\bibnamefont
  {Ilderton}}\ and\ \bibinfo {author} {\bibfnamefont {G.}~\bibnamefont
  {Torgrimsson}},\ }\href {\doibase 10.1016/j.physletb.2013.07.045} {\bibfield
  {journal} {\bibinfo  {journal} {Phys. Lett. B}\ }\textbf {\bibinfo {volume}
  {725}},\ \bibinfo {pages} {481} (\bibinfo {year} {2013})}\BibitemShut
  {NoStop}%
\bibitem [{\citenamefont {Tamburini}\ \emph {et~al.}(2010)\citenamefont
  {Tamburini}, \citenamefont {Pegoraro}, \citenamefont {Piazza}, \citenamefont
  {Keitel},\ and\ \citenamefont {Macchi}}]{Tamburini:2010}%
  \BibitemOpen
  \bibfield  {author} {\bibinfo {author} {\bibfnamefont {M.}~\bibnamefont
  {Tamburini}}, \bibinfo {author} {\bibfnamefont {F.}~\bibnamefont {Pegoraro}},
  \bibinfo {author} {\bibfnamefont {A.~D.}\ \bibnamefont {Piazza}}, \bibinfo
  {author} {\bibfnamefont {C.~H.}\ \bibnamefont {Keitel}}, \ and\ \bibinfo
  {author} {\bibfnamefont {A.}~\bibnamefont {Macchi}},\ }\href
  {http://stacks.iop.org/1367-2630/12/i=12/a=123005} {\bibfield  {journal}
  {\bibinfo  {journal} {New Journal of Physics}\ }\textbf {\bibinfo {volume}
  {12}},\ \bibinfo {pages} {123005} (\bibinfo {year} {2010})}\BibitemShut
  {NoStop}%
\bibitem [{\citenamefont {{J. D. Jackson}}(1999)}]{jackson}%
  \BibitemOpen
  \bibfield  {author} {\bibinfo {author} {\bibnamefont {{J. D. Jackson}}},\
  }\href@noop {} {\emph {\bibinfo {title} {{Classical Electrodynamics}}}}\
  (\bibinfo  {publisher} {John Wiley \& Sons, Inc., New York},\ \bibinfo {year}
  {1999})\BibitemShut {NoStop}%
\bibitem [{\citenamefont {Harvey}\ \emph {et~al.}(2011)\citenamefont {Harvey},
  \citenamefont {Heinzl},\ and\ \citenamefont {Marklund}}]{Harvey:2011dp}%
  \BibitemOpen
  \bibfield  {author} {\bibinfo {author} {\bibfnamefont {C.}~\bibnamefont
  {Harvey}}, \bibinfo {author} {\bibfnamefont {T.}~\bibnamefont {Heinzl}}, \
  and\ \bibinfo {author} {\bibfnamefont {M.}~\bibnamefont {Marklund}},\ }\href
  {\doibase 10.1103/PhysRevD.84.116005} {\bibfield  {journal} {\bibinfo
  {journal} {Phys. Rev. D}\ }\textbf {\bibinfo {volume} {84}},\ \bibinfo
  {pages} {116005} (\bibinfo {year} {2011})}\BibitemShut {NoStop}%
\bibitem [{\citenamefont {Di~Piazza}\ \emph {et~al.}(2009)\citenamefont
  {Di~Piazza}, \citenamefont {Hatsagortsyan},\ and\ \citenamefont
  {Keitel}}]{DiPiazza:2009RR}%
  \BibitemOpen
  \bibfield  {author} {\bibinfo {author} {\bibfnamefont {A.}~\bibnamefont
  {Di~Piazza}}, \bibinfo {author} {\bibfnamefont {K.~Z.}\ \bibnamefont
  {Hatsagortsyan}}, \ and\ \bibinfo {author} {\bibfnamefont {C.~H.}\
  \bibnamefont {Keitel}},\ }\href {\doibase 10.1103/PhysRevLett.102.254802}
  {\bibfield  {journal} {\bibinfo  {journal} {Phys. Rev. Lett.}\ }\textbf
  {\bibinfo {volume} {102}},\ \bibinfo {pages} {254802} (\bibinfo {year}
  {2009})}\BibitemShut {NoStop}%
\bibitem [{\citenamefont {Harvey}\ \emph {et~al.}(2009)\citenamefont {Harvey},
  \citenamefont {Heinzl},\ and\ \citenamefont {Ilderton}}]{PhysRevA.79.063407}%
  \BibitemOpen
  \bibfield  {author} {\bibinfo {author} {\bibfnamefont {C.}~\bibnamefont
  {Harvey}}, \bibinfo {author} {\bibfnamefont {T.}~\bibnamefont {Heinzl}}, \
  and\ \bibinfo {author} {\bibfnamefont {A.}~\bibnamefont {Ilderton}},\ }\href
  {\doibase 10.1103/PhysRevA.79.063407} {\bibfield  {journal} {\bibinfo
  {journal} {Phys. Rev. A}\ }\textbf {\bibinfo {volume} {79}},\ \bibinfo
  {pages} {063407} (\bibinfo {year} {2009})}\BibitemShut {NoStop}%
\bibitem [{\citenamefont {Esarey}\ \emph {et~al.}(1993)\citenamefont {Esarey},
  \citenamefont {Ride},\ and\ \citenamefont {Sprangle}}]{Esarey:1993}%
  \BibitemOpen
  \bibfield  {author} {\bibinfo {author} {\bibfnamefont {E.}~\bibnamefont
  {Esarey}}, \bibinfo {author} {\bibfnamefont {S.~K.}\ \bibnamefont {Ride}}, \
  and\ \bibinfo {author} {\bibfnamefont {P.}~\bibnamefont {Sprangle}},\ }\href
  {\doibase 10.1103/PhysRevE.48.3003} {\bibfield  {journal} {\bibinfo
  {journal} {Phys. Rev. E}\ }\textbf {\bibinfo {volume} {48}},\ \bibinfo
  {pages} {3003} (\bibinfo {year} {1993})}\BibitemShut {NoStop}%
\bibitem [{\citenamefont {Heinzl}\ \emph {et~al.}(2010)\citenamefont {Heinzl},
  \citenamefont {Seipt},\ and\ \citenamefont {K\"ampfer}}]{Seipt:2012}%
  \BibitemOpen
  \bibfield  {author} {\bibinfo {author} {\bibfnamefont {T.}~\bibnamefont
  {Heinzl}}, \bibinfo {author} {\bibfnamefont {D.}~\bibnamefont {Seipt}}, \
  and\ \bibinfo {author} {\bibfnamefont {B.}~\bibnamefont {K\"ampfer}},\ }\href
  {\doibase 10.1103/PhysRevA.81.022125} {\bibfield  {journal} {\bibinfo
  {journal} {Phys. Rev. A}\ }\textbf {\bibinfo {volume} {81}},\ \bibinfo
  {pages} {022125} (\bibinfo {year} {2010})}\BibitemShut {NoStop}%
\bibitem [{\citenamefont {Boca}\ and\ \citenamefont
  {Florescu}(2009)}]{Boca:2009}%
  \BibitemOpen
  \bibfield  {author} {\bibinfo {author} {\bibfnamefont {M.}~\bibnamefont
  {Boca}}\ and\ \bibinfo {author} {\bibfnamefont {V.}~\bibnamefont
  {Florescu}},\ }\href {\doibase 10.1103/PhysRevA.80.053403} {\bibfield
  {journal} {\bibinfo  {journal} {Phys. Rev. A}\ }\textbf {\bibinfo {volume}
  {80}},\ \bibinfo {pages} {053403} (\bibinfo {year} {2009})}\BibitemShut
  {NoStop}%
\bibitem [{\citenamefont {Harvey}\ \emph {et~al.}(2012)\citenamefont {Harvey},
  \citenamefont {Heinzl}, \citenamefont {Ilderton},\ and\ \citenamefont
  {Marklund}}]{Harvey:2012ie}%
  \BibitemOpen
  \bibfield  {author} {\bibinfo {author} {\bibfnamefont {C.}~\bibnamefont
  {Harvey}}, \bibinfo {author} {\bibfnamefont {T.}~\bibnamefont {Heinzl}},
  \bibinfo {author} {\bibfnamefont {A.}~\bibnamefont {Ilderton}}, \ and\
  \bibinfo {author} {\bibfnamefont {M.}~\bibnamefont {Marklund}},\ }\href
  {\doibase 10.1103/PhysRevLett.109.100402} {\bibfield  {journal} {\bibinfo
  {journal} {Phys. Rev. Lett.}\ }\textbf {\bibinfo {volume} {109}},\ \bibinfo
  {pages} {100402} (\bibinfo {year} {2012})}\BibitemShut {NoStop}%
\bibitem [{\citenamefont {Ritus}(1985)}]{Ritus:1985}%
  \BibitemOpen
  \bibfield  {author} {\bibinfo {author} {\bibfnamefont {V.}~\bibnamefont
  {Ritus}},\ }\href {\doibase 10.1007/BF01120220} {\bibfield  {journal}
  {\bibinfo  {journal} {Journal of Soviet Laser Research}\ }\textbf {\bibinfo
  {volume} {6}},\ \bibinfo {pages} {497} (\bibinfo {year} {1985})}\BibitemShut
  {NoStop}%
\bibitem [{\citenamefont {Sauter}(1931)}]{QEDcriticalfield1}%
  \BibitemOpen
  \bibfield  {author} {\bibinfo {author} {\bibfnamefont {F.}~\bibnamefont
  {Sauter}},\ }\href {\doibase 10.1007/BF01339461} {\bibfield  {journal}
  {\bibinfo  {journal} {Z.~Phys.}\ }\textbf {\bibinfo {volume} {69}},\ \bibinfo
  {pages} {742} (\bibinfo {year} {1931})}\BibitemShut {NoStop}%
\bibitem [{\citenamefont {Heisenberg}\ and\ \citenamefont
  {Euler}(1936)}]{QEDcriticalfield2}%
  \BibitemOpen
  \bibfield  {author} {\bibinfo {author} {\bibfnamefont {W.}~\bibnamefont
  {Heisenberg}}\ and\ \bibinfo {author} {\bibfnamefont {H.}~\bibnamefont
  {Euler}},\ }\href {\doibase 10.1007/BF01343663} {\bibfield  {journal}
  {\bibinfo  {journal} {Z.~Phys.}\ }\textbf {\bibinfo {volume} {98}},\ \bibinfo
  {pages} {714} (\bibinfo {year} {1936})}\BibitemShut {NoStop}%
\bibitem [{\citenamefont {Schwinger}(1951)}]{QEDcriticalfield3}%
  \BibitemOpen
  \bibfield  {author} {\bibinfo {author} {\bibfnamefont {J.}~\bibnamefont
  {Schwinger}},\ }\href {\doibase 10.1103/PhysRev.82.664} {\bibfield  {journal}
  {\bibinfo  {journal} {Phys. Rev.}\ }\textbf {\bibinfo {volume} {82}},\
  \bibinfo {pages} {664} (\bibinfo {year} {1951})}\BibitemShut {NoStop}%
\bibitem [{\citenamefont {Harvey}\ \emph {et~al.}(2015)\citenamefont {Harvey},
  \citenamefont {Marklund},\ and\ \citenamefont {Wallin}}]{Harvey:2015rca}%
  \BibitemOpen
  \bibfield  {author} {\bibinfo {author} {\bibfnamefont {C.}~\bibnamefont
  {Harvey}}, \bibinfo {author} {\bibfnamefont {M.}~\bibnamefont {Marklund}}, \
  and\ \bibinfo {author} {\bibfnamefont {E.}~\bibnamefont {Wallin}},\
  }\href@noop {} {\  (\bibinfo {year} {2015})},\ \Eprint
  {http://arxiv.org/abs/1507.06478} {arXiv:1507.06478 [hep-ph]} \BibitemShut
  {NoStop}%
\end{thebibliography}
%

\end{document}